# Conformal Shield: A Novel Adversarial Attack Detection Framework for Automatic Modulation Classification

Tailai Wen, Da Ke, Xiang Wang, and Zhitao Huang

*Abstract*—Deep learning algorithms have become an essential component in the field of cognitive radio, especially playing a pivotal role in automatic modulation classification. However, Deep learning also present risks and vulnerabilities. Despite their outstanding classification performance, they exhibit fragility when confronted with meticulously crafted adversarial examples, posing potential risks to the reliability of modulation recognition results. Addressing this issue, this letter pioneers the development of an intelligent modulation classification framework based on conformal theory, named the Conformal Shield, aimed at detecting the presence of adversarial examples in unknown signals and assessing the reliability of recognition results. Utilizing conformal mapping from statistical learning theory, introduces a custom-designed Inconsistency Soft-solution Set, enabling multiple validity assessments of the recognition outcomes. Experimental results demonstrate that the Conformal Shield maintains robust detection performance against a variety of typical adversarial sample attacks in the received signals under different perturbation-to-signal power ratio conditions.

## I. INTRODUCTION

Automatic Modulation Classification (AMC) technology is an integral part of both cooperative and non-cooperative communication scenarios in future communication systems. In cooperative communication scenarios, the most typical application is cognitive Internet of Things (IoT). In densely populated IoT environments, where spectrum resources are limited, AMC can aid IoT systems in more efficiently utilizing the available spectrum, enhancing overall communication efficiency [1-2]. Furthermore, IoT devices often employ diverse communication technologies and standards. AMC enables these IoT devices to adaptively comprehend and communicate using different modulation schemes, ensuring compatibility between devices and reducing communication interference [3]. In non-cooperative communication scenarios, the most typical application is Cognitive Electronic Warfare (CEW), where AMC is an important tool for accurately identifying and classifying enemy communication and radar signals, providing vital intelligence for electronic countermeasures.

This capability allows for precise targeting and disruption of enemy systems, while minimizing interference with friendly communications. Additionally, AMC enhances the adaptability of CEW systems, enabling them to effectively respond to rapidly changing electronic warfare environments and various signal types [4],[5].

In previous research, deep learning algorithms have nearly perfected the accuracy of AMC techniques, integrating various feature processing technologies to accommodate modulation recognition in different conditions, such as low signal-to-noise ratios [6]. However, deep learning methods are susceptible to a type of carefully designed, imperceptible perturbation to the human eye, known as adversarial attacks, which can cause wildly incorrect classification results. As demonstrated in Figure 1, unlike traditional methods of interference, these attacks are more covert and specifically target intelligent models. If an attacker can perfectly intercept a signal from the sender, generate a perturbation, and then synchronize and relay it into the radio frequency spectrum, as depicted in Figure 2, the original receiver would receive a signal sample contaminated with the adversarial attack. This would severely disrupt the backend intelligent modulation classification mechanism, causing different devices to receive incorrect modulations and potentially crippling the entire electronic system.

Whether in cooperative or non-cooperative communication scenarios, the initial process for an attacker is to prevent the receiver from restoring the signal from the transmitter. Therefore, when the attacker detects communication activity in a frequency band, preventing successful AMC becomes the primary goal [7]. For the receiver, the biggest challenge is to detect whether the signal sample has been attacked. Failure to promptly discover errors in modulation type classification will lead to continuously matching incorrect decoding schemes, consuming a significant amount of time and resources to solve. In the face of the emerging attack strategies targeting AMC, although existing researches [8-10] have developed defenses against modulation classification attacks, these defense measures often rely on the premise of "knowing how the opponent will pose the problem" to provide "open-book"

This work of Tailai Wen, Xiang Wang and Zhitao Huang was supported by the Natural Science Foundation of China (Grant No. 62271494). This work of Tailai Wen has been supported by the Hunan Provincial Postgraduate Research Innovation Programme under Grant CX20230043.
Tailai Wen, Da Ke, Xiang Wang and Zhitao Huang are with State Key Laboratory of Complex Electromagnetic Environment Effects on Electronics and Information System, College of Electronic Science and Technology, National University of Defense Technology, Changsha 410073, China (e-mail:wentailai@nudt.edu.cn;1747884404@qq.com;christopherwx@163.com; huangzhitao@nudt.edu.cn).

answers. They base their defense criterion on how to ensure the accuracy of AMC does not decrease in the presence of adversarial samples. In reality, attackers often manage to discern the type and parameters of the target model through means akin to 'espionage' activities. However, the defense side struggles to accurately predict the specific attack strategy that will be employed. Therefore, the timely detection of attacks on the received signals and the real-time assessment of the effectiveness of AMC results become exceedingly important.

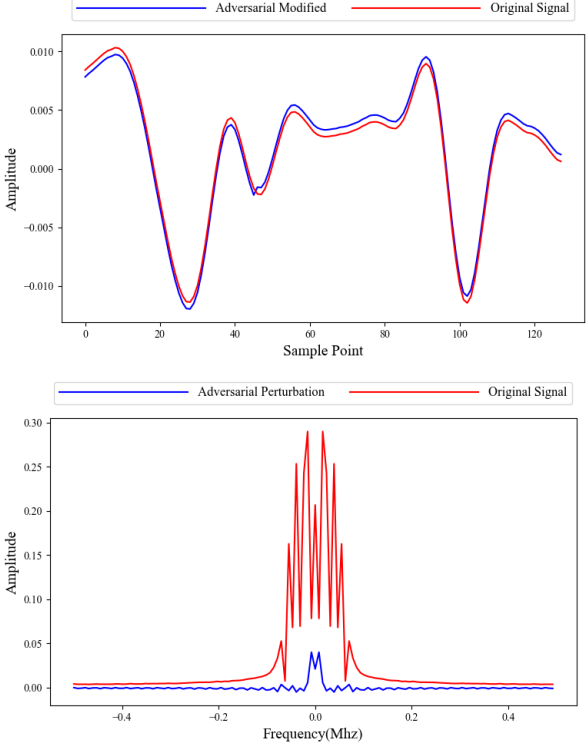

**Fig. 1.** Time-domain waveform and Spectrum of a communication signal and the adversarial perturbation added.

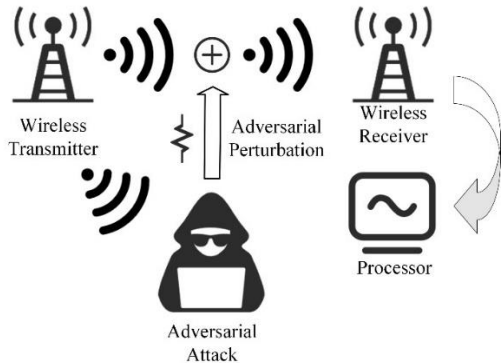

**Fig. 2.** A ideal scenario for the adversarial examples in modulation classification.

To address this issue, we first consider the worst-case attack scenario for the defender, where the attacker has stolen the parameters of our recognition model and can create carefully designed adversarial samples to synchronize the contaminated signal samples into the channel through the transmitter. Then, on the defense side, inspired by conformal theory [11], we designed the Conformal Shield under the original intelligent recognition model and defined an Inconsistency Soft-solution Set (ISS) to detect whether the signal sample has been attacked. Overall, the main contributions of this paper are as follows:

1. We are the first to introduce conformal theory into the defense framework against adversarial problems in wireless modulation signal classification.

2. The Conformal Shield technology we propose is a strategy completely different from previous defense perspectives, assessing the reliability of recognition results from a universal perspective. We verify the advanced nature of the Conformal Shield by conducting experiments with different attack methods in different conditions.

## II. THE PROPOSED FRAMEWORK TO DETECT ADVERSARIAL ATTACK

### A. Adversarial attack in signals

We commence by delineating the methodology for generating adversarial perturbations aimed at signal modulation recognition models. In the domain of AMC that leverages deep learning networks as its foundational architecture, I/Q (in-phase/quadrature) plots of signals under varying Signal-to-Noise Power Ratio (SNR) conditions are conventionally employed as input samples, with the type of modulation utilized as labels for the purpose of training, eventually facilitating the AMC of unknown signals [6]. Drawing inspiration from the realm of adversarial examples in computer vision, [7] has integrated the Fast Gradient Sign Method (FGSM) into AMC, effectuating a successful assault on the model and thereby precipitating a diminution in the accuracy of recognition. Pertaining to a classifier built upon Deep Neural Networks (DNN), the fabrication of adversarial samples can be articulated as an optimization problem encumbered with constraints as follow:

$$\max_\delta L_f(\theta, x+\delta, y)$$
$$\text{s.t. } \|\delta\|_\infty \leq \varepsilon \text{ and } f(x) \neq f(x+\delta) \quad (1)$$

Where $L_f$ represents the loss function of the DNN in AMC. $\delta$ is a perturbation doped into the signal. $\varepsilon$ is the small positive constant, and the smaller the value the more deceptive the $\delta$ obtained. $y$ represents the real label corresponding to the input signal I/Q sample $x$. For the FGSM, $\delta$ is solved as follows when only a one-step attack is considered:

$$\delta = \varepsilon \cdot \text{sign}(\nabla_x L_f(\theta, x+\delta, y)) \quad (2)$$

Where $\nabla$ represent a gradient descent. In addition, the more precise and potent attack methodology known as Projected Gradient Descent (PGD) has been introduced into the realm of signal classification. This approach meticulously refines an initially generated random perturbation through an iterative process. The iterative equation is as follows:

$$x'_t = x + \delta$$
$$x'_{t+1} = \text{ProD}\left(x'_t + \beta \cdot sign(\nabla_x L_f(\theta, x'_t, y))\right) \quad (3)$$

ProD stands for projection operation to ensure that the perturbed sample is within the allowed range. $\beta$ represents the step size of the perturbation during each iteration. Furthermore, there exists another methodology known as Carlini & Wagner (CW), which achieves more covert attacks by introducing finer regularization terms. This method stands as one of the most representative techniques in the generation of adversarial examples. CW can also be implemented within the framework of PGD, and its form of implementation is as follows:

$$x'_{t+1} = \text{ProD}_{L_2}\left(x'_t + \beta \cdot sign\left(\nabla_x L_f\left(\theta, x'_t, y\right) + \lambda \cdot \left\|x'_t - x\right\|_2\right)\right) \quad (4)$$

We utilize Residual Network (Resnet) as the underlying recognition model to construct the AMC mechanism. The generation of adversarial examples as a means of attack is carried out employing the aforementioned three methods. The relative strength of the perturbations is quantified using the Perturbation-to-Signal Power Ratio (PSR), as delineated in [7]. The expression for this metric is as follows:

$$\text{PSR} = 10\lg\left(P_\delta / P_y\right) \quad (5)$$

*B. Conformal prediction*

As discussed in the previous section, in response to adversarial attacks, we have formulated two hypotheses: 1) Adversarial examples, in their endeavor to maintain stealth, for DNN models predicated on probabilistic predictions, elevate the probabilities of alternative classes post-attack, thus disrupting the typical probability distribution characterized by a high probability for the correct class and low probabilities for other classes, without necessarily resulting in a distribution where an incorrect class dominates. 2) When the output of a DNN model is no longer a singular label but presented in the form of a set, reliability assessment becomes feasible through the examination of the output. Consequently, we have decided to incorporate Conformal Prediction (CP), transforming the probability predictor into the set predictor, in an attempt to validate these hypotheses.

This section primarily focuses on describing the variant of CP, known as Split CP (S-CP), which is notably more applicable for handling large datasets and scenarios requiring efficient computation. Assume the presence of a signal sample set $D$, which is partitioned based on a pre-defined ratio into a training set $D^{tr}$ with $N^{tr}$ samples and a validation set $D^{val}$ comprising $N^{val}$ samples. The error level $\alpha$ of the prediction set is predetermined. For any probabilistic predictor $p(y|x,D^{tr})$ that is to be trained, the samples from the training set are used. Subsequently, in the testing phase, for a given unknown signal sample $x$, the final prediction set $S(x|D)$, determining which type of modulation ($y' \in Y$) it contains, is deduced based on the "conformity" of the sample pair $(x,y')$ with the validation set.

In the testing phase, for each sample pair, 'conformity' can be quantified using a Nonconformity Score (NCS). In the context of Split CP (S-CP), this NCS can be articulated using the logarithmic loss of the model, as follows:

$$\text{NCS}(z=(x,y)|D^{tr}) = -logp(y|x,D^{tr}) \quad (6)$$

Alternatively, distinct integral metrics may be utilized to assess the loss incurred by the probabilistic predictor $p(y|x,D^{tr})$ during the evaluation of unknown samples. Subsequent to the computation of the NCS utilizing the identical model across diverse modulation labels, the ultimate aggregation of the sample set, wherein the sample is posited, is ascertained in accordance with the predetermined level of $\alpha$. From a mathematical standpoint, the set predictor in the S-CP framework can be derived via the ensuing formulation:

$$S(x|D) = \left\{ \begin{array}{l} y' \in Y | \text{ NCS}\left((x,y')|D^{tr}\right) \\ \leq Q_\alpha\left(\left\{\text{NCS}\left((x_i,y_i)^{val}|D^{tr}\right)\right\}_{i=1}^{N^{val}}\right) \end{array} \right\} \quad (7)$$

Where $Q_\alpha\left(\{z[i]\}_{i=1}^N\right)$ represent the $\lceil(1-\alpha)(N+1)\rceil$-th smallest value of the set $\{z[i]\}_{i=1}^N \cup \{+\infty\}$.

Utilizing S-CP, we are able to transform any DNN model into a set predictor. Inspired by this theoretical framework, we further innovate to create the Conformal Shield paradigm.

*C. Conformal Shield*

In order to validate Hypothesis 1 posited in the previous section, and to directly monitor whether unknown signals are subject to attack, this section will introduce the Conformal Shield Framework that we have pioneered.

To enhance the robustness of the S-CP model, it is first transformed into a CP under the K-fold cross-validation, followed by the integration of our own defined ISS, forming a complete conformal shield framework. The detailed steps are illustrated in Figure 3.

We divide the dataset $D$ (total number of N samples) into $K$ ($K \geq 2$) subsets $\{S_k\}_{k=1}^{k=K}$. During the training phase, $K$-1 subsets are used for training, yielding $K$ probability predictors $p(y|x,S_k)$. For each $p(y|x,S_k)$, the subset $S_k$ (contains $Z_k$ pairs of samples) not used for training is utilized to calculate $p(y|x,S_k)$'s NCS set $\text{NCS}(Z_k|D-S_k)$. In the testing phase, for a given test signal sample $x$ and its possible corresponding label $y'$ ($y' \in Y$), the $\text{NCS}((x,y')|D-S_k)$ corresponding to the unknown signal is calculated for all K $p(y|x,S_k)$ s. Then, $\text{NCS}((x,y')|D-S_k)$ is ranked against each $p(y|x,S_k)$'s NCS set and the total number N(y') of all NCSs in the NCS sets of all K-fold $p(y|x,S_k)$ s that are larger than $\text{NCS}((x,y')|D-S_k)$ is counted. If N(y') is larger than $\alpha$ percent of N, then the label $y'$ can be included in the

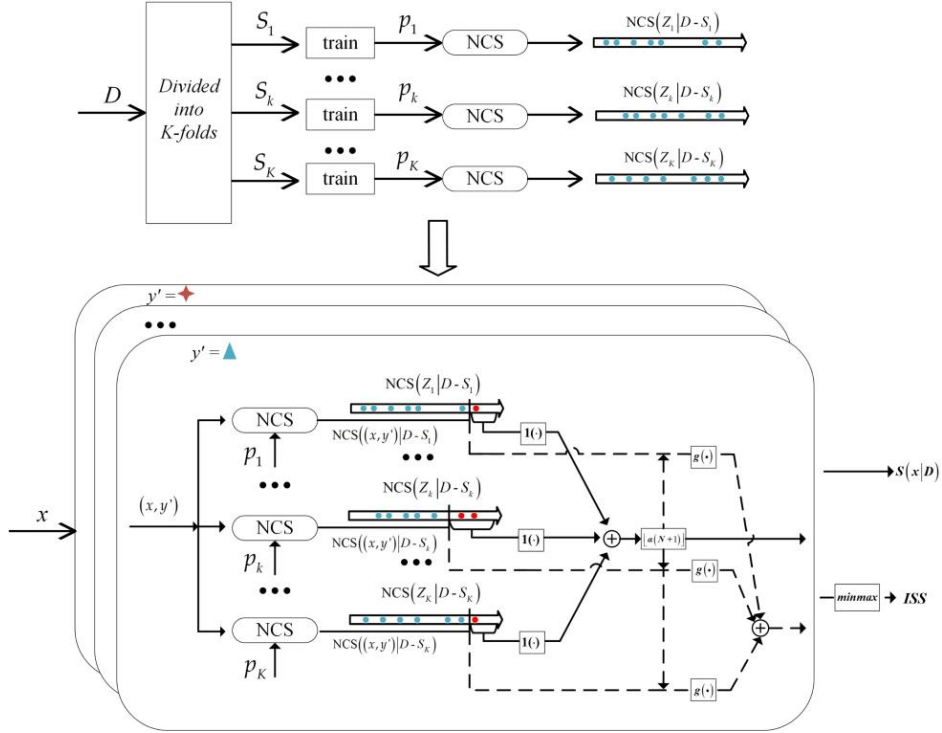

**Fig. 3.** Flamework of Conformal Shield

final prediction set. The mathematical expression of the prediction set is as follows:

$$S(x|D) = \left\{ y' \in Y \mid \sum_{k=1}^{K} \sum_{Z_k \in S_k} 1(hi) \geq \lfloor \alpha(N+1) \rfloor \right\}$$
$$hi = \text{NCS}((x,y')| D - S_k) \leq \text{NCS}(Z_k | D - S_k) \quad (8)$$

Where $1(\cdot)$ represents the hard decision function. The metrics commonly used to assess the performance of the set predictor are coverage and inefficiency, expressed as follows:

$$cover(S) = \text{Pro}(y \in S(x|D)) \geq 1-\alpha \quad (9)$$

$$ineffi(S) = \text{E}\left[|S(x|D)|\right] \quad (10)$$

Considering the real-world scenario, where the samples inputted to AMC are typically signal samples of a certain modulation sliced by duration, we designed an ISS, combined with inefficiency, to jointly monitor the adversarial nature of the prediction set. We refer to this framework as the Conformal Shield. The mathematical expression of the ISS is as follows:

$$ISS = minmax\left( y' \in Y / \sum_{k=1}^{K} \sum_{x \in X} g\left(\text{NCS}((x,y')| si)\right) \right) \quad (11)$$
$$si = 1(hi) \geq \lfloor \alpha(N+1) \rfloor$$

The term $g(\cdot)$ represents a function that reassigns values less than 0 to 0, while $X$ denotes the entire signal segment to which the sample $x$ belongs. The function $minmax$ indicates the normalization of element values to the interval [0, 1].

Ultimately, by integrating inefficiency and the ISS, we examine the reliability of unknown signal recognition outcomes. This constitutes the essence of the Conformal Shield.

## III. RESULTS AND DISCUSSION

In this section, we will employ the three attack methodologies mentioned in Section II to conduct experimental assaults on the AMC model. And it is to validate the detection performance of the Conformal Shield. The dataset is sourced from the GNU radio ML dataset RML2016.19a, published in [6] and available at https://www.deepsig.ai/datasets. It encompasses 11 modulation signals: BPSK, QPSK, 8PSK, QAM16, QAM64, CPFSK, GFSK, PAM4, WBFM, AM-SSB, and AM-DSB. Each modulation type comprises 20 different SNR levels, ranging from -20 dB to 18 dB, with increments of 2 dB per level. For every SNR level, each modulation signal includes 1100 signal samples, with each sample containing 128-point complex I/Q channel symbols.

Selecting signals with an SNR greater than or equal to 10dB (comprising a total of 55,000 signal samples), 80% of these are used for training the Resnet, while the remaining 20% serve as unknown signals for testing phase. During the division of the training and test datasets, stratified proportional sampling is employed to ensure that the test samples maintain the same distribution as the training samples. To ensure the stealth of adversarial perturbations, the PSR range is fixed between -20 dB and 0 dB for testing, and the unknown signals used for input are exclusively those with an SNR level of 16 dB. In the Conformal Shield, $\alpha$ is set to 0.1, and K to 10. During the testing phase, only modulation type signals with high recognition rates were selected for the attack experiment.

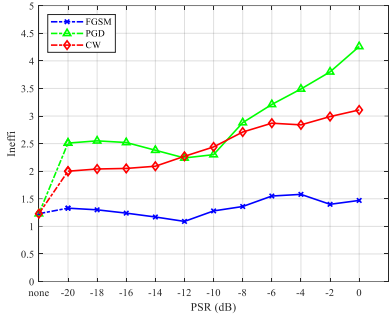

**Fig. 4.** Results of the inefficacy by three attack approaches across varied PSR level

Tab. 1 Results of the ISS by three attack approaches with PSR=-20dB

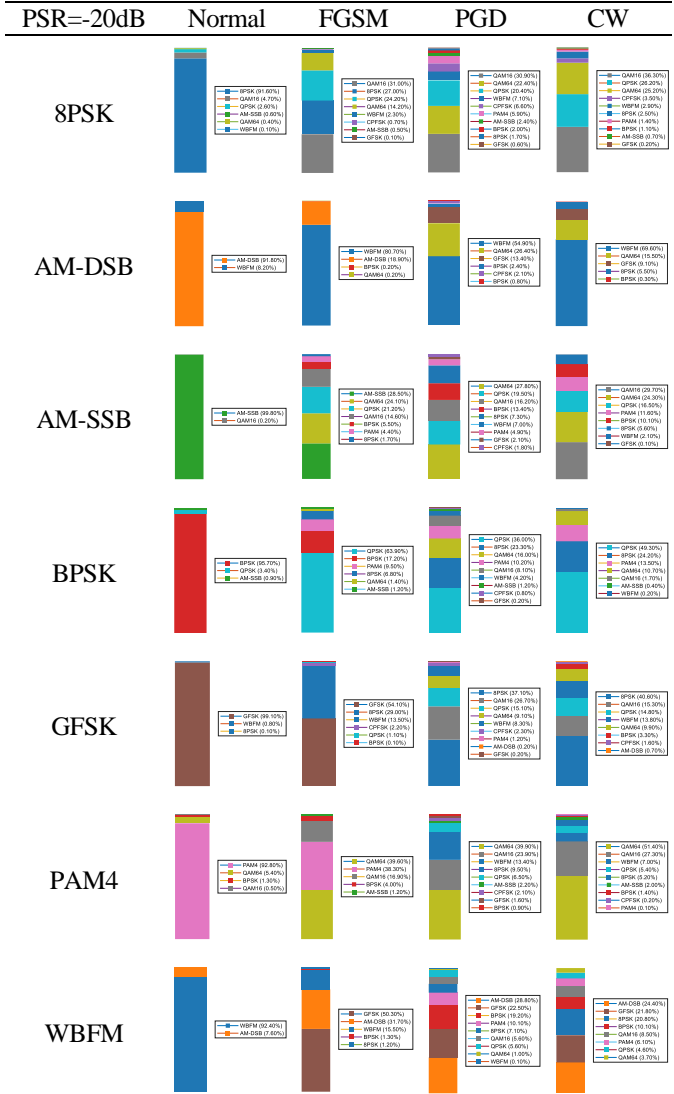

In Figure 4, for strong attack methods like PGD and CW, inefficacy significantly increases with PSR. However, for weaker attack methods such as FGSM and in cases of lower PSR level, it is challenging to demonstrate the reliability of the $S(x|D)$. Yet, when analyzing in conjunction with the ISS, as illustrated in Table 1, it is evident that even at PSR = -20 dB, there is a noticeable shift in the probability distribution of the prediction results for the sliced samples of the received signal. This change does not fulfill the conformal theory condition of the correct category probability distribution being greater than $\alpha$. The results indicate that the Conformal Shield framework proposed in this paper can detect adversarial attacks on the received signals through ISS analysis.

## IV. CONCLUSION

We are the first to propose the Conformal Shield framework for detecting adversarial attacks in AMC. Experiments on AMC subjected to three types of adversarial attacks (FGSM, PGD and CW) were conducted using a publicly accessible modulated signal dataset. Notably, the Conformal Shield is capable of assessing the reliability of AMC outcomes even in scenarios with low perturbation energy level. Our future work involves further refining the Conformal Shield framework, such that in scenarios of weak attack modes, the recognition results can maintain a high coverage rate. In cases of strong attack modes, the system will be enhanced to more effectively identify the nature of the attack and its intentions, such as determining whether there is a presence of targeted inducement.